\documentclass[preprint, 12pt]{elsarticle}

\usepackage{amssymb}
\usepackage{amsmath}
\usepackage{mathtools}

\usepackage{amsthm}
\newtheorem{proposition}{Proposition}[subsection]
\newtheorem{problem}{Problem}[subsection]

\usepackage{lineno}
\usepackage{xcolor}
\usepackage{float}

\journal{Physica D (Nonlinear Phenomena)}

\newcommand{\prj}{\mathcal{P}}
\newcommand{\R}{\mathbb{R}}
\newcommand{\E}{\mathbb{E}}

\begin{document}

\begin{frontmatter}

\title{Ensemble of Fixed Points in Multi-branch Shell Models of Turbulent Cascades}

\author[label1]{Flavio Tuteri}
\ead{flavio.tuteri@phys.ens.psl.eu}

\author[label2]{Sergio Chibbaro}
\author[label1]{Alexandros Alexakis}

\affiliation[label1]{
    organization={Laboratoire de Physique de l’École normale supérieure, ENS, Université PSL, CNRS, Sorbonne Université, Université Paris Cité},
    city={Paris},
    postcode={F-75005},
    country={France}
}

\affiliation[label2]{
    organization={Laboratoire Interdisciplinaire des Sciences du Numérique, LISN, Université Paris-Saclay, CNRS, CentraleSupélec, Inria},
    city={Orsay},
    postcode={F-91405},
    country={France}
}


\begin{abstract}
Stationary solutions of a shell model of turbulence defined on a dyadic tree topology are studied. 
Each node's amplitude is expressed as the product of amplitude multipliers associated with its
ancestors, providing a recursive representation of the cascade process.
A geometrical rule governs the tree growth, and we prove the existence of a continuum of fixed points—including the Kolmogorov solution—that sustain a strictly forward energy cascade.
Sampling along randomly chosen branches defines a homogeneous Markov chain, enabling a stochastic characterization of extended self-similarity and intermittency through the spectral properties of the associated Feynman–Kac operators.
Numerical simulations confirm the theoretical predictions, showing that multi-branch shell models offer a minimal yet physically rich framework for exploring the complexity of nonlinear energy transfer across scales.
\end{abstract}



\end{frontmatter}



\section{Introduction}
\label{sec1}

\begin{figure}[t]
\centering
\includegraphics[width=\columnwidth]{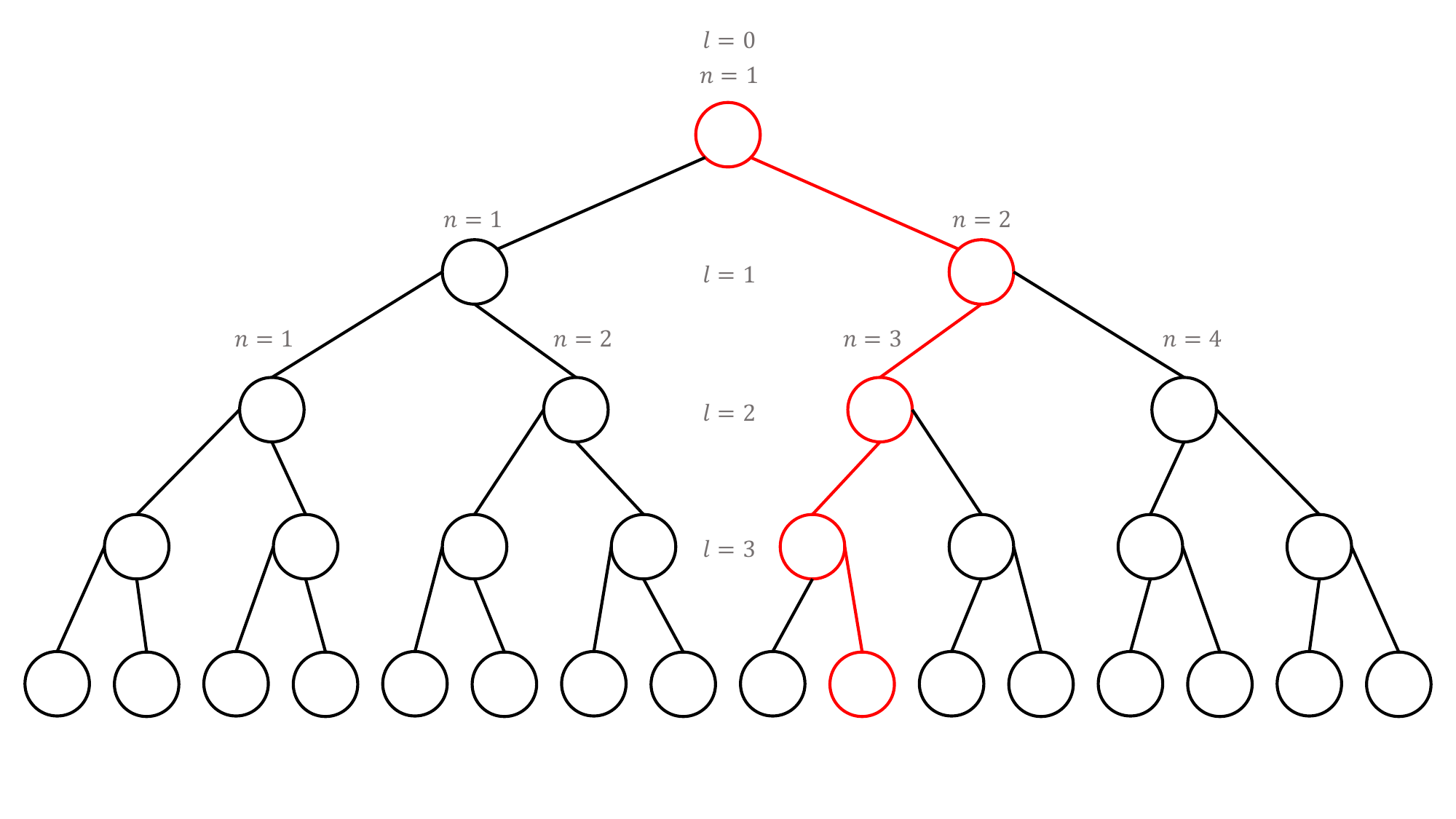}
\caption{Schematic representation of the $2$-adic tree with one descent path highlighted.}
\label{figTree}
\end{figure}

Explaining the phenomenology of conserved-quantity transfer across scales in turbulent flows remains a central problem in non-equilibrium physics \cite{alexakis2018}. Richardson’s cascade picture \cite{richardson2007} describes energy percolation from large to progressively smaller eddies. Tracing this behavior directly from the Navier–Stokes equations is challenging, as nonlinear advection couples all scales while the pressure field, enforcing incompressibility, couples all spatial locations. Shell models \cite{biferale2003} provide a tractable framework that captures the hallmark features of three-dimensional turbulence—forward mean energy flux, non-Gaussian statistics, and chaotic dynamics—while remaining numerically accessible.\medskip

Classical single-branch models encode the cascade as a linear hierarchy across scales. They are heuristically derived from a coarse-graining of Fourier space into logarithmically spaced shells, each associated with a single dynamical variable representing the velocity amplitude of the given scale. The dynamical variables are then coupled, retaining only the interactions between nearest neighbors, consistently to \cite{eyink2005}.
This formulation, however, flattens the geometrical and topological structure of the flow, restricting any intermittency to be found in the temporal behavior of the dynamical variables.\medskip

Observations of turbulent flows reveal spatially localized and intermittent energy transfer, hinting at a more complex branching organization. Embedding shell dynamics within a multi-branch topology, like the one illustrated in Figure~\ref{figTree}, provides a unified framework for exploring both spatial intermittency \cite{frisch1995} —associated with Richardson-like splitting— and temporal intermittency observed in standard shell chains \cite{aumaitre2024,mailybaev2013}.
Early numerical investigations of shell models defined on such trees \cite{benzi1997,aurell1994,aurell1997} were hindered by the rapidly increasing computational cost, as the number of modes grows exponentially when smaller scales are resolved.
More recently, an ensemble of exact stationary solutions has been constructed in this topology \cite{ajzner2023}, leveraging the nearest-neighbour interactions of the Desnianskii–Novikov model \cite{desnianskii1974}. These solutions make it possible to probe arbitrarily small scales and demonstrate that the model can exhibit intermittent behavior.
Furthermore, the multi-branch approach connects naturally with multiresolution wavelet analysis \cite{benzi1997}; indeed, using the wavelet transform, one can heuristically justify this class of hierarchical models from the fluid equations.
\medskip

Here, we introduce and analyze a shell model with GOY-type couplings \cite{gledzer1973,obukhov1971,yamada1987} defined on a $p$-adic tree: starting from a root, each level is generated by assigning $p$ children to every node. Exploiting a geometrical construction, we build an ensemble of stationary solutions that sustain a strictly forward cascade. Our main theoretical result is an explicit existence proof for a continuum of such states—including the Kolmogorov solution—showing inductively that any admissible configuration up to a given level can always be extended to the next. Remarkably, the multi-branch case gives rise to infinitely many fixed points, in contrast to the single-branch case. Moreover, this ensemble is rich and includes intermittent solutions.\medskip

Beyond existence, we consider a sampling procedure along randomly chosen branches, yielding a homogeneous Markov chain. Adopting the viewpoint of stochastic dynamical systems allows for the analysis of self-similarity and intermittency through ergodic properties \cite{hairer2011} and spectral analysis \cite{hennion2001} of Feynman–Kac operators, providing a route to scaling exponents that accounts for interlevel correlations. As pointed out in \cite{biferale1994,mailybaev2023}, operator analysis proves to be a powerful framework for studying the scaling properties of nonlinear systems.
Numerical experiments implementing the geometric sampler confirm the theoretical predictions. \medskip

The results establish multi-branch shell models, sparsely studied in the literature \cite{benzi1997,aurell1994,aurell1997,ajzner2023}, as a minimal yet analytically controllable setting in which forward local energy transfer and strong intermittent statistics coexist.


\section{Foundations}
\label{sec2}

This section introduces a shell model of turbulence defined on a tree topology and analyzes the existence of stationary solutions under physically motivated constraints.


\begin{figure}[t]
\centering
\includegraphics[width=\columnwidth]{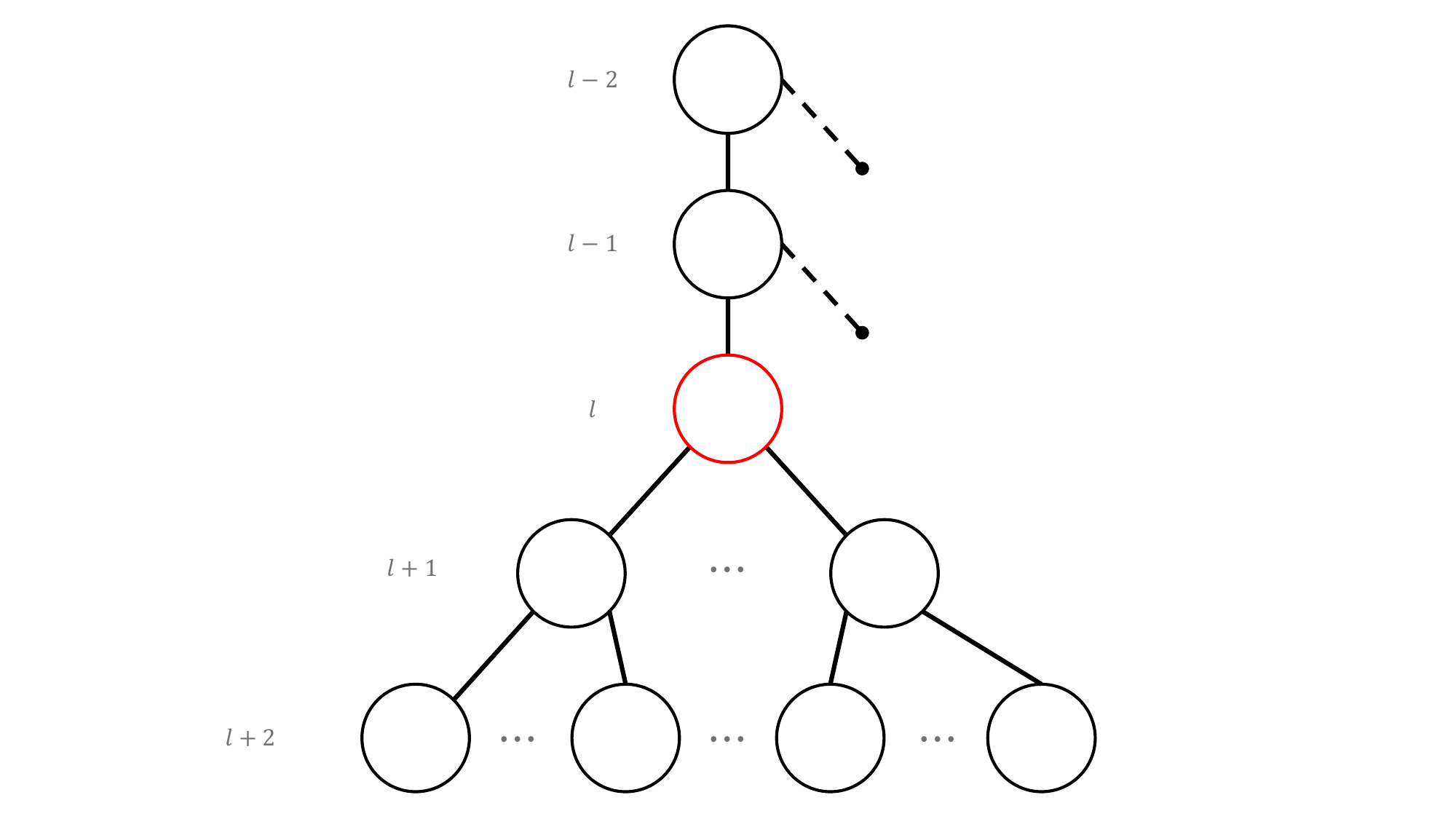}
\caption{Schematic of the nonlinear interactions affecting the red-highlighted node, showing contributions from first and second neighbors in the GOY-like model.}
\label{figInteract}
\end{figure}

\subsection{System definition}

Fix an intershell ratio $\lambda>1$ and define the shell wavenumbers as $k_l=\lambda^l$.
The system is defined on a $p$-adic tree (see Figure \ref{figTree} for the case $p=2$), with node set
\begin{equation}
    \big\{ (l,n);\,l=0,\ldots,l_{max},\,n = 1,\ldots,p^l \big\}.
\end{equation}
Each node $(l,n)$, except the root $(0,1)$, has a single parent $(l-1,\lceil n/p\rceil)$, where $\lceil x \rceil = \min \{ n \in \mathbb{Z} \mid n \ge x \}$ is the ceiling function. All non-terminal nodes are connected to their $p$ children $(l+1,p(n-1)+1),\ldots,(l+1,pn)$.
The GOY model on this topology is written as
\begin{equation}
    \bigg[\frac{d}{dt}+\nu{k_l}^{2}\bigg]u_{l,n}=i N_{l,n}[u]^\ast+f_{l,n},
\end{equation}
where $\nu$ is the viscosity, $f$ the large-scale forcing, 
and the nonlinear coupling is given by
\begin{alignat}{3}
    N_{l,n}=k_{l-1}\bigg[&a\,\lambda\,\frac{1}{p}\sum_{h=0}^{p-1}\bigg( &&u_{l+1,pn-h}               \,\frac{1}{p}\sum_{m=0}^{p-1} &&u_{l+2,p(pn-h)-m}\bigg)\nonumber\\
                      +\,&b                                       &&u_{l-1,\lceil n/p\rceil}   \,\frac{1}{p}\sum_{m=0}^{p-1} &&u_{l+1,pn-m}            \nonumber\\
                      +\,&c\,\lambda^{-1}                         &&u_{l-2,\lceil n/p^2\rceil}                               &&u_{l-1,\lceil n/p\rceil}\bigg],
\label{nonlinear}
\end{alignat}
with boundary conditions $u_{-2,\bullet}=u_{-1,\bullet}=0=u_{l_{max}+1,\bullet}=u_{l_{max}+2,\bullet}$. Figure \ref{figInteract} clarifies the structure of the nonlinear interactions: all GOY-like combinations obtained by varying the horizontal indices contribute equally.
The inviscid and unforced dynamics conserves the total energy, defined as a volume-weighted sum of the local energy density,
\begin{equation}
    E=\frac{1}{2}\sum_{l=0}^{l_{max}}\frac{1}{{k_l}^D}\sum_{n=1}^{p^l}\lvert u_{l,n}\rvert^2,
\end{equation}
provided that
\begin{equation}
    a+b+c=0,
\label{energy_conservation}
\end{equation}
and
\begin{equation}
    D^{-1}=\log_p\lambda,
\label{spaDim}
\end{equation}
where $D$ denotes the effective spatial dimension. Relation (\ref{spaDim}) has a clear physical meaning: for a space-filling cascade, the number of substructures $p$ equals the volume reduction factor between successive scales.
Observables are defined in accordance with the interpretation of $l$ as a scale index and $n$ as a spatial index.
To ensure, in addition, the conservation of a helicity-like invariant,
\begin{equation}
    H=\sum_{l=0}^{l_{max}}(-1)^l{k_l}^{1-D}\sum_{n=1}^{p^l}\lvert u_{l,n}\rvert^2,
\end{equation}
we choose 
\begin{equation}
    a=1,\quad b=-(1-\lambda^{-1}),\quad c=-\lambda^{-1}.
\label{model_parameters}
\end{equation}
The local energy transfer across an edge (i.e., across the branch that connects $(l-1,\lceil n/p\rceil)$ with $(l,n)$) is defined as the rate of change of the energy contained in $(l,n)$ and in all descendant nodes:
\begin{alignat}{2}
    \Pi_{l,n}&=\operatorname{Im}\!\bigg[\sum_{j=l}^{l_{max}}&&\frac{1}{{k_j}^D}\sum_{m=p^{j-l}(n-1)+1}^{p^{j-l}n} u_{j,m}N_{j,m}[u]\bigg] \\
             &=\frac{k_{l-1}}{{k_l}^D}\,\operatorname{Im}\!\bigg[&&-a\,u_{l-1,\lceil n/p\rceil}u_{l,n}\,\frac{1}{p}\sum_{m=0}^{p-1}u_{l+1,pn-m} \nonumber\\
             &&&+c\,\lambda^{-1}\, u_{l-2,\lceil n/p^2\rceil}u_{l-1,\lceil n/p\rceil}u_{l,n}\bigg].
\label{local_flux}
\end{alignat}
We have derived expression (\ref{local_flux}) using energy conservation (\ref{energy_conservation}), extending the algebra used for classical shell models \cite{pisarenko1993}. Hereafter, we restrict to the dyadic case ($p=2$).


\subsection{Problem statement}

We investigate stationary states of the inviscid system, characterized by the condition $N_{l,n}[u]=0$. Since the forcing is active at large scales, it can be incorporated into the boundary conditions for the nonlinear term.
Assuming non-vanishing shell amplitudes, we introduce the normalized velocity multipliers
\begin{equation}
    r_{l,n}=\lambda^{1/3}\frac{u_{l,n}}{u_{l-1,\lceil n/2\rceil}},\qquad l\ge1.
\end{equation}
The nonlinear dynamics defined by (\ref{nonlinear}) can be recast as a recurrence relation involving four consecutive levels:
\begin{align}
    \frac{a}{4}\Big[{r_{l+1,2n-1}}^2(r_{l+2,4n-3}+r_{l+2,4n-2})&+{r_{l+1,2n}}^2(r_{l+2,4n-1}+r_{l+2,4n})\Big]\nonumber\\
    +\frac{b}{2}\,{r_{l,n}}^{-1}(r_{l+1,2n-1}+r_{l+1,2n})&+c\,{r_{l-1,\lceil n/2\rceil}}^{-1}{r_{l,n}}^{-2}=0.
\label{recurrence}
\end{align} 
This can be considered a recursion relation in $l$: given the tree up to level $l+1$, it is possible to construct level $l+2$. The shell variables can then be reconstructed as products of these multipliers:
\begin{equation}
    u_{l,n}=\lambda^{-l/3}\,u_{0}\prod_{j=0}^{l-1}r_{l-j,\lceil n/2^j\rceil}.
\end{equation}
Owing to the quadratic homogeneity of the nonlinear coupling, 
the system is invariant under uniform rescaling of the amplitudes. 
We therefore fix the global phase and normalization by setting $u_{0}=i$. 
With this convention, and considering real-valued multipliers, the local energy transfer (\ref{local_flux}) becomes
\begin{equation}
    \Pi_{l,n}=\frac{k_{l-1}}{2^l}(-iu_{l,n})^3\bigg[\frac{a}{2}\,{r_{l,n}}^{-1}(r_{l+1,2n-1}+r_{l+1,2n})-c\,{r_{l-1,\lceil n/2\rceil}}^{-1}{r_{l,n}}^{-2}\bigg].
\label{new_flux}
\end{equation}
We assume that the local energy flux is non-vanishing, 
so that the right flux multipliers can be introduced as
\begin{align}
    \pi_{l+1,2n}&=\frac{\Pi_{l+1,2n}}{\Pi_{l,n}}=\frac{\Pi_{l+1,2n}}{(-iu_{l+1,2n})^3}\, \frac{{r_{l+1,2n}}^3}{\lambda}\, \frac{(-iu_{l,n})^3}{\Pi_{l,n}}\nonumber\\
    &=\frac{1}{2}\,{r_{l+1,2n}}^2\,\frac{a(r_{l+2,4n-1}+r_{l+2,4n})/2-c\,{r_{l,n}}^{-1}{r_{l+1,2n}}^{-1}}{a\,{r_{l,n}}^{-1}(r_{l+1,2n-1}+r_{l+1,2n})/2-c\,{r_{l-1,\lceil n/2\rceil}}^{-1}{r_{l,n}}^{-2}}
\label{flux_ratios}
\end{align}
and, analogously, the left flux multipliers, $\pi_{l+1,2n-1}$.
By constraining the flux multipliers to be positive, the local energy flux maintains the same sign throughout the tree; its sign depends on the initial conditions.
Given this geometrical formulation, all multipliers $r_{l,n}$ can be found recursively, provided that allowed solutions of (\ref{recurrence}) do not vanish and that the condition $\pi_{l,n}>0 $ is not violated. Since we consider stationary solutions, the latter case would correspond to the unphysical situation of an energy source at smaller scales. We would then like to show that there is a bounded set $I\subset\R$ such that if all $r_{j,\bullet}\in I$  for $j\le l+1$ then the set of solutions $(r_{l+2,4n-3},r_{l+2,4n-2},r_{l+2,4n-1},r_{l+2,4n})$ of (\ref{recurrence}) that also satisfy the flux inequality has a non-empty intersection with $I$. If this holds, one can always find solutions of (\ref{recurrence}) recursively and construct the entire tree geometry. More precisely, the problem can be formulated as follows.
\begin{problem}
\label{problem}
    We seek stationary solutions of (\ref{recurrence}) that remain bounded within the interval $I=[-\xi,-\eta]\cup[\eta,\xi]$, with $\xi>1>\eta>0$, thus ensuring regularity and including the Kolmogorov fixed point $r\equiv1$ within the admissible ensemble. A physically relevant stationary state is further required to sustain a positive local energy flux, corresponding to a direct transfer of energy toward smaller scales and consistent with the absence of energy sources at small scales. 
    From the geometrical perspective, an inductive construction is required: for every configuration up to a given level, it must be possible to iterate the recurrence within the desired subregion defined by all imposed constraints.
\end{problem}


\subsection{Geometrical interpretation}
\label{exres}

Whenever unambiguous, indices are omitted for brevity.
The recurrence relation (\ref{recurrence}) admits a natural geometric interpretation in $\mathbb{R}^4$. We construct
\begin{equation}
    \mathbf{x}=(r_{l+2,4n-3},r_{l+2,4n-2},r_{l+2,4n-1},r_{l+2,4n})
\end{equation}
constrained to lie on the affine hyperplane $A$ defined by $\langle\boldsymbol{\alpha},\mathbf{x}\rangle=\beta$, which is exactly (\ref{recurrence}) for
\begin{align}
    \boldsymbol{\alpha} &= \frac{a}{4} \Big({r_{l+1,2n-1}}^2,{r_{l+1,2n-1}}^2,{r_{l+1,2n  }}^2,{r_{l+1,2n  }}^2\Big),\\
    \beta &= -b\,{r_{l  ,n   }}^{-1}\frac{1}{2}(r_{l+1,2n-1}+r_{l+1,2n  })
             -c\,{r_{l  ,n   }}^{-2}\,         {r_{l-1,\lceil n/2\rceil}}^{-1}.
\end{align}
Parameters $\boldsymbol{\alpha}$ and $\beta$ are determined by the ancestor multipliers.
The uniform boundedness condition confines the admissible configurations to the region $I^4$.
Flux positivity, expressed through the flux multipliers (\ref{flux_ratios}), imposes an additional geometric constraint: the admissible configurations must lie within the intersection of two half-spaces defined by
\begin{equation}
    \sigma\bigg[\frac{1}{2}(x_1+x_2)-\psi_1\bigg]>0,\qquad\sigma\bigg[\frac{1}{2}(x_3+x_4)-\psi_2\bigg]>0,
\label{half_spaces}
\end{equation}
where $\sigma$ is the sign of
\begin{equation}
    \phi=\frac{a}{2}\,{r_{l,n}}^{-1}(r_{l+1,2n-1}+r_{l+1,2n})-c\,{r_{l-1,\lceil n/2\rceil}}^{-1}{r_{l,n}}^{-2},
\end{equation}
which is the denominator in (\ref{flux_ratios}), and
\begin{equation}
    \psi_1=\frac{c}{a}\,{r_{l,n}}^{-1}{r_{l+1,2n-1}}^{-1},\qquad\psi_2=\frac{c}{a}\,{r_{l,n}}^{-1}{r_{l+1,2n}}^{-1}.
\end{equation}
We proceed by deriving a parametrization for the hyperplane $A$. The affine orthogonal projector onto $A$ is
\begin{equation}
    \prj\mathbf{x}=\mathbf{x}-\frac{\langle\boldsymbol{\alpha},\mathbf{x}\rangle-\beta}{{\lVert\boldsymbol{\alpha}\rVert}^2}\,\boldsymbol{\alpha}.
\end{equation}
We now note that the tangent space $\{\boldsymbol{\alpha}=(\alpha_1,\alpha_1,\alpha_2,\alpha_2)\}^\perp$ has the orthogonal basis
\begin{align}
    \mathbf{b}_1=(1,-1, 0, 0),\quad
    \mathbf{b}_2=(0, 0, 1,-1),\quad\mathbf{b}_3=(\alpha_2,\alpha_2,-\alpha_1,-\alpha_1).
\end{align}
Hence, any point in $A$ can be expressed as
\begin{equation}
    \mathbf{x}=\prj\mathbf{0}+\sum_{i=1}^3 t_i\mathbf{b}_i
\label{param}
\end{equation}
with $\mathbf{t}\in\R^3$. 
Thus, upon substitution, the flux-positivity condition (\ref{half_spaces}) becomes
\begin{equation}
    L_0\vcentcolon=\frac{\sigma}{\alpha_2}\bigg[\psi_1-\frac{\beta}{\lVert\boldsymbol{\alpha}\rVert^2}\alpha_1\bigg]<\sigma t_3<\frac{\sigma}{\alpha_1}\bigg[\frac{\beta}{\lVert\boldsymbol{\alpha}\rVert^2}\alpha_2-\psi_2\bigg]\vcentcolon=U_0.
\label{t3_interval}
\end{equation}
This interval is always non-empty, since
\begin{equation}
    U_0-L_0=\sigma\bigg[\frac{\beta}{\lVert\boldsymbol{\alpha}\rVert^2}\bigg(\frac{\alpha_1}{\alpha_2}+\frac{\alpha_2}{\alpha_1}\bigg)-\frac{1}{\alpha_1\alpha_2}(\alpha_1\psi_1+\alpha_2\psi_2)\bigg]=\frac{\lvert\phi\rvert}{2\alpha_1\alpha_2}>0,
\label{nonempty}
\end{equation}
where for the last equality we have used
\begin{equation}
    \beta-\phi=2(\alpha_1\psi_1+\alpha_2\psi_2),
\end{equation}
which follows from the energy conservation condition (\ref{energy_conservation}). Finally, we state our fundamental result, concerning the construction of an infinite tree as an inductive solution to Problem \ref{problem}.
\begin{proposition}
\label{thm}
    Given $\lambda>1$, for any $\eta\in(0,1)$ and any $\xi>\xi_0(\lambda,\eta)$, where
    \begin{equation}
        \xi_0=\max\bigg\{3\eta,\frac{1}{\eta^2}\bigg[\frac{1}{\lambda\eta^3}+\bigg(1-\frac{1}{\lambda}\bigg)\bigg]\bigg\},
    \end{equation}
    there exists a continuum of admissible iterations of the recurrence relation (\ref{recurrence}) that satisfy both the boundedness condition in $I$ and the flux-positivity constraints (\ref{half_spaces}).
\end{proposition}
The proof is purely geometric and is provided in \ref{app1}. We remark that letting $\eta$ approach zero requires increasingly large values of $\xi$. These parameters therefore control the admissible fluctuations of the resulting solutions: a larger allowed interval $I$ leads to the possibility of more intermittent behaviors.


\section{Sampling}
\label{sec3}

In the previous section, we proved that one can construct a continuum of stationary solutions iteratively using (\ref{recurrence}) by picking (randomly or deterministically) real parameters $t_1$, $t_2$ and $t_3$ within their allowed limits. In what follows, we show that these solutions can be complex enough to display intermittent scaling, as observed in solutions of the Navier–Stokes equations. We begin by introducing a framework for the theoretical analysis and numerical exploration of these stationary solutions.


\subsection{Multiplicative random walk}
\label{samp_process}

\begin{figure}[!h]
\centering
\includegraphics[width=\columnwidth]{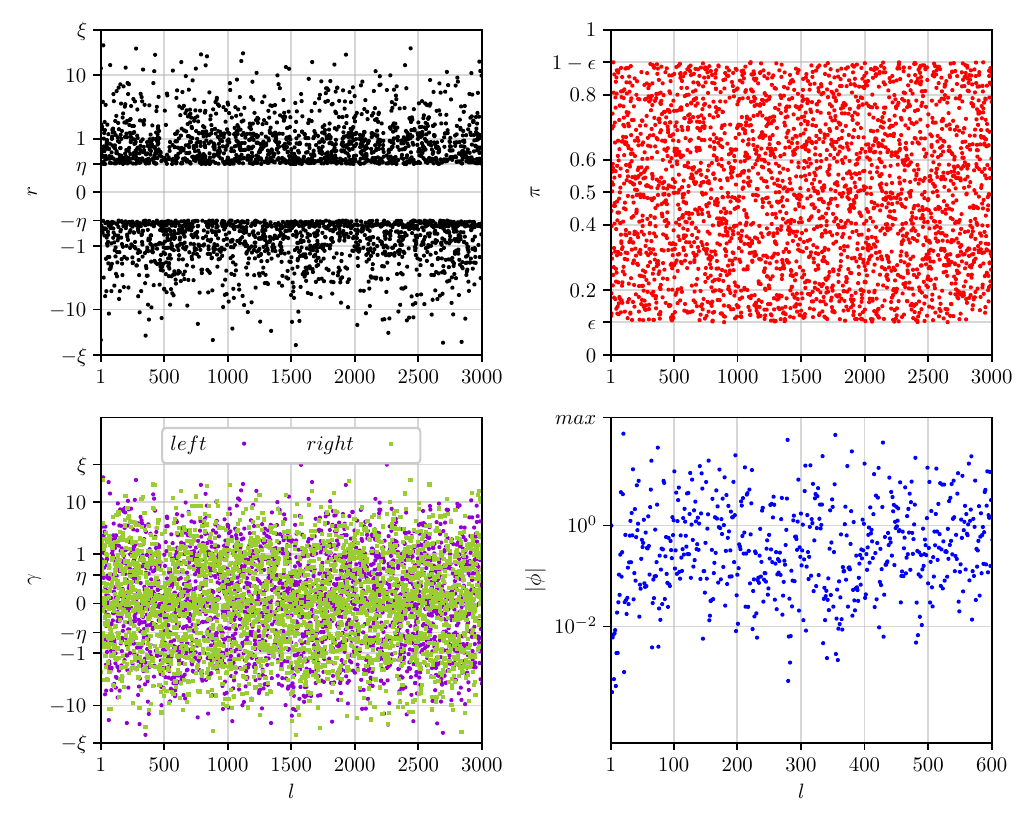}
\caption{Simulation of $3000$ steps of the Markov chain.
The top-left panel shows the evolution of the amplitude multiplier, demonstrating its confinement within the prescribed region.
The top-right panel displays the trajectory of the flux multipliers, bounded between $\epsilon$ and $1-\epsilon$.
The bottom panels illustrate the remaining geometric functions relevant to the construction. In particular, the right inset shows a zoomed portion for $\lvert\phi\rvert$, highlighting the absence of patterns associated with extended self-similarity.}
\label{figTraj}
\end{figure}

We consider a randomly chosen descent of the tree, like the one illustrated in red in Figure \ref{figTree}, parametrized by $n_l$, giving the horizontal index $n$ of that path at level $l$. This process is a Markov chain of order $3$, because step $l+2$ depends on steps $l+1$, $l$, and $l-1$, but we can extend the state to define a standard order $1$ homogeneous Markov chain on the state space $\Omega = I^4$:
\begin{equation}
    Z_l=\left(r_{l-1,\lceil n_l/2\rceil},\,r_{l,n_l},\,r_{l+1,2n_l-1},\,r_{l+1,2n_l}\right).
\label{chainState}
\end{equation}
The chain evolves via a rejection-sampling update. Let $(L,U)$ denote the final admissible interval for $t_3$, enforcing
\begin{equation}
    \pi_{l+1,2n-1}>\epsilon, \qquad\pi_{l+1,2n}>\epsilon,
\end{equation}
with $\epsilon\in(0,1/2)$. This additional constraint, useful in simulations to limit the ensemble size, is discussed in \ref{app1}.
At each step, the new state is generated according to (\ref{param}) as follows. First,
\begin{equation}
    t_3\text{ is drawn uniformly from }(L,U),
\label{samp1}
\end{equation}
thus forming $\boldsymbol{\gamma}=\prj\mathbf{0}+t_3\mathbf{b}_3=(\gamma_1,\gamma_1,\gamma_2,\gamma_2)$. Then,
\begin{align}
    \lvert t_i\rvert\text{ is drawn uniformly from }&\lvert\gamma_i\rvert+\eta+[0,\tau_1]\text{ if }\lvert\gamma_i\rvert\le\eta,\nonumber\\
    \text{ and from }&[0,\tau_2]\text{ otherwise},
\label{samp2}
\end{align}
for $i=1,2$. Samples falling within the forbidden strips $\{|x_i|<\eta\}$ are rejected. The index of the next node $n_{l+1}$ is selected as $2n_l-1$ or $2n_l$ with equal probability, corresponding to a fair coin toss.
An example realization of this process is illustrated in Figure \ref{figTraj}, showing velocity and flux multipliers, the intermediate stage $\boldsymbol{\gamma}$ for the new guess of amplitude multipliers, and $\phi$, a measure of the ensemble size. \medskip 

Energy conservation along this process ensures
\begin{equation}
    \E[2\pi_{l+1,n_{l+1}}\vert Z_l]=\E[\pi_{l+1,2n_l-1}+\pi_{l+1,2n_l}\vert Z_l]=1,
\end{equation}
so that, to compute the expectation of the flux, the martingale property can be used:
\begin{align}
    \E\big[2^l\Pi_{l,n}\big]&=\E\big[\E\big[2^l\Pi_{l,n}\vert Z_{l-1}\big]\big]=\E\big[\E\big[2^{l-1}\Pi_{l-1,\lceil n/2\rceil}2\pi_{l,n}\vert Z_{l-1}\big]\big] \nonumber\\
    &=\E\big[2^{l-1}\Pi_{l-1,\lceil n/2\rceil}\E\big[2\pi_{l,n}\vert Z_{l-1}\big]\big]=\E\big[2^{l-1}\Pi_{l-1,\lceil n/2\rceil}\big]=C.
\label{prediction}
\end{align}
The constant $C$ does not depend on $l$ and represents the mean energy injection rate.
We have multiplied the local flux by the number of nodes in that level, which is exactly what is needed to reconstruct the ensemble average, given the equivalence with respect to the spatial average, i.e. the horizontal mean. \medskip

The sampling process given in (\ref{samp1}), (\ref{samp2}) defines a transition kernel
\begin{equation}
    K(\mathbf{z},B) = \mathbb{P}(Z_{l+1}\in B \mid Z_l=\mathbf{z}),
\end{equation}
where $B\subset\Omega$ is a Borel set, i.e. $K(\mathbf{z},B)$ is the probability of $Z_{l+1}\in B$ given that $Z_l=\mathbf{z}$. This kernel governs the multiplicative random walk along the tree.
Then, the probability of $Z_l\in B$ given the initial condition $Z_0=\mathbf{z}$ is the $l$-times iterated kernel $K^l$ given by
\begin{align}
    \mathbb{P}(Z_{l}\in B\vert Z_{0}=\mathbf{z})&=
    \int_\Omega d\mathbf{w}_{1}\,\mathbb{P}(Z_{1}=\mathbf{w}_{1}\vert Z_{0}=\mathbf{z})\cdots\nonumber\\
    \cdots\int_\Omega &d\mathbf{w}_{l-1}\,\mathbb{P}(Z_{l-1}=\mathbf{w}_{l-1}\vert Z_{l-2}=\mathbf{w}_{l-2})
    \mathbb{P}(Z_{l}\in B\vert Z_{l-1}=\mathbf{w}_{l-1})\nonumber\\
    &=\int_\Omega K(\mathbf{z},d\mathbf{w}_{1})\cdots\int_\Omega K(\mathbf{w}_{l-2},d\mathbf{w}_{l-1})K(\mathbf{w}_{l-1},B). \label{Kl}
\end{align} 
Equation (\ref{Kl}) provides all the information about the statistics of $Z_l$ at any level $l$. Its asymptotic behavior as $l\to\infty$ is examined in the next subsection.


\subsection{Ergodicity and intermittency}

The asymptotic behavior of $K^l$ for large $l$ is revealed by ergodic theorems for Markov chains, such as Theorem $1.2$ in \cite{hairer2011}, which ensures the existence of a unique invariant measure $\mu_\infty$ satisfying
\begin{equation}
\label{ergo}
    \lVert K^l-\mu_\infty\rVert\xrightarrow{l\to\infty} 0,\qquad\text{for any initial state }\mathbf{z}\in\Omega,
\end{equation}
implying that the iterated kernel becomes independent of $Z_0$ as $l\to \infty$ and that the distributions of the multipliers become independent of $l$, i.e. self-similar. 
\medskip

However, in order to quantify intermittency, we need moments of $u_{l,n}$ or $\Pi_{l,n}$, which amounts to computing moments of products of multipliers.
Suppose that we are interested in the $p$-th moment of an observable $O$, expressed as a function $g_O$ of $Z_l$ and $Z_{l+1}$.
For this purpose, we introduce the Feynman–Kac operators acting on continuous functions $f$ over $\Omega$ as
\begin{equation}
    [\mathcal{J}_p f](\mathbf{z})=\int_\Omega \vert g_O(\mathbf{z},\mathbf{z}^\prime)\vert^p f(\mathbf{z}^\prime)K(\mathbf{z},d\mathbf{z}^\prime),
\end{equation}
so that 
\begin{equation}
    \mathbb{E}[\lvert O(Z_l,Z_{l+1})\rvert^p|Z_{l}={\bf z}] = [\mathcal{J}_p \mathbf{1}](\mathbf{z}).
\label{link}
\end{equation}
For the velocity multipliers $r_{l,n_l}$, the corresponding function $g_r$ is simply
\begin{equation}
    g_r(\mathbf{z},\mathbf{z}^\prime)={z_2}^\prime,
\end{equation}
while for the energy flux we need $2\pi_{l,n_l}$, obtained by defining $g_\pi$ as
\begin{align}
    g_\pi(\mathbf{z},\mathbf{z}^\prime)=\frac{z_2/2}{a(z_3+z_4)/2-c{z_1}^{-1}{z_2}^{-1}}&\nonumber\\
    \times\,\bigg\{{z_3}^2\Big[a(z^\prime_1+z^\prime_2)/2-c{z_2}^{-1}{z_3}^{-1}\Big]
    +&\,{z_4}^2\Big[a(z^\prime_3+z^\prime_4)/2-c{z_2}^{-1}{z_4}^{-1}\Big]\bigg\}. 
\end{align}

The Feynman–Kac operators defined here allow us to exploit spectral theory. 
The link to the quantity of interest, here being the local energy flux times the volume, is given by equation (\ref{link}). By iterating this operator, we recover the products of multipliers:
\begin{equation}
    \mathbb{E}\bigg[\prod_{j=2}^{l+1}\vert 2\pi_{j, n_j}\vert^p\bigg]=\mathbb{E}\big[{\mathcal{J}_p}^l \mathbf{1}\big].
\end{equation}
Statistical correlations among levels are accounted for in the Feynman–Kac operator, since it is defined in terms of the transition kernel.
Invoking the asymptotic spectral decomposition \cite{hennion2001} for large $l$,
\begin{equation}
\label{asym}
    {\mathcal{J}_p}^lf=\mathcal{R}(\mathcal{J}_p)^l\bigg\{\varphi_p\bigg[\int_\Omega f\, d\nu_p\bigg]+o(1)\bigg\},
\end{equation}
where $\nu_p$ is a probability measure on $\Omega$ (bra, in Quantum Mechanics jargon), $\varphi_p$ is a continuous positive function on $\Omega$ (ket), normalized so that $\int_\Omega \varphi_p\,d\nu_p=1$, and $\mathcal{R}$ denotes the spectral radius, that is, the supremum of the magnitudes of the elements of the spectrum. Thus,
\begin{equation}
    \mathbb{E}\big[{\mathcal{J}_p}^l \mathbf{1}\big]\sim \mathbb{E}[\varphi_p]\,\mathcal{R}(\mathcal{J}_p)^l\propto\lambda^{-l\tau_p},
\end{equation}
with
\begin{equation}
    \tau_p=-\log_\lambda \mathcal{R}(\mathcal{J}_p).
\end{equation}
The result above shows that the moments of the observable $\Pi_{l,n_l}$ follow a power-law scaling with $l$.
Furthermore, since in general $\mathcal{R}(\mathcal{J}_p) \ne 1$, the exponents $\tau_p$ are not all zero; in other words, the solutions exhibit intermittent scaling.
We note that this scaling law holds despite the presence of complex inter-level correlations in our model. 
Although we are able to justify a power law, the above expression for the exponents $\tau_p$ is not easy to compute numerically.
In the following section, we will compute the scaling directly.


\section{Simulations}
\label{sec4}
\begin{figure}[h]
\centering
\includegraphics[width=\columnwidth]{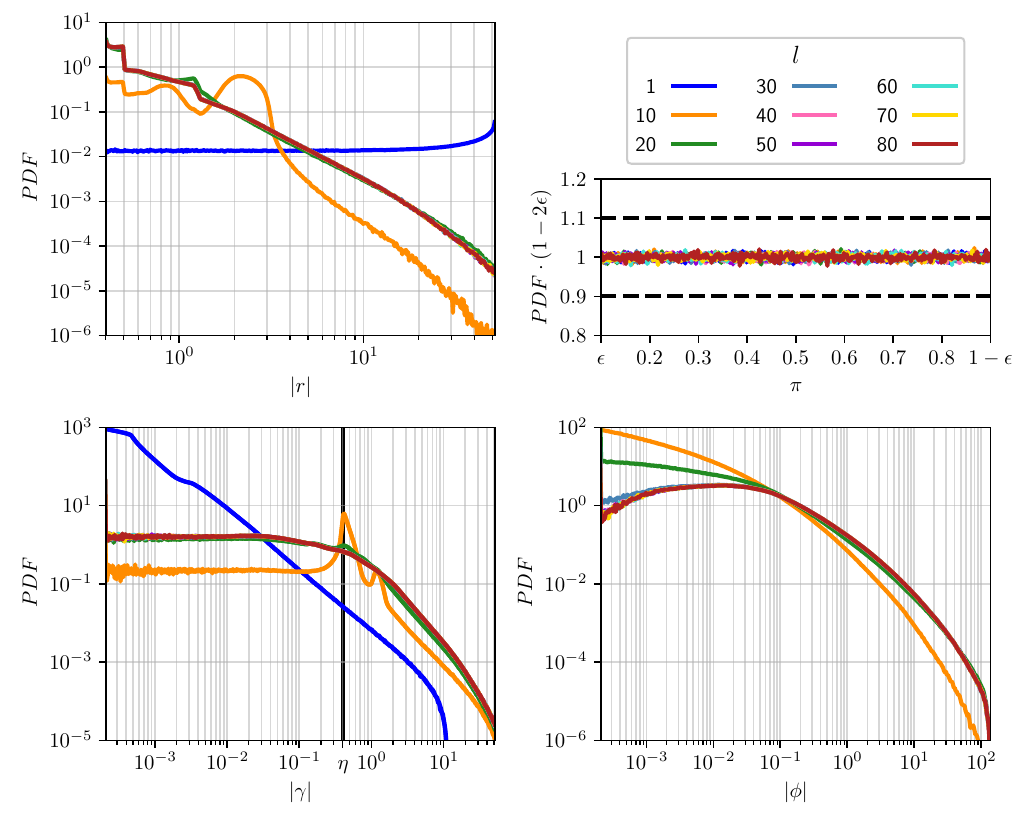}
\caption{Probability distributions of the multipliers and geometric functions defined in the sampling process.
Results are obtained from $10^7$ samples, showing collapse for sufficiently large $l$.
In the top-right panel, a uniform distribution is observed for the flux multipliers;
the dashed lines denote a $10\%$ deviation from uniform probability over the support.}
\label{figHist}
\end{figure}
We performed Monte Carlo simulations of the multiplicative random walk described in Section \ref{samp_process}. In particular, we considered the case $\lambda=2$, which corresponds to $D=1$ as in \cite{benzi1997}. 
Figure \ref{figTraj} shows the results for a $3000$-step run with parameters $\epsilon=0.1$, $\tau_1=\tau_2=0.1$, $\eta=0.4$, and $\xi=52$. The confinement of the multipliers is verified for this long trajectory. Moreover, the absence of patterns in $l$ for $|\phi|$ is significant, as it relates to the extended self-similarity property.
This follows from the relation
\begin{equation}
    \lambda\,2^l\Pi_{l,n}=k_{l}\lvert u_{l,n}\rvert^3\,\lvert\phi_{l,n}\rvert
\end{equation}
which arises directly from the definitions and from the positivity of the flux. 
\medskip
\begin{figure}[t]
\centering
\includegraphics[width=\columnwidth]{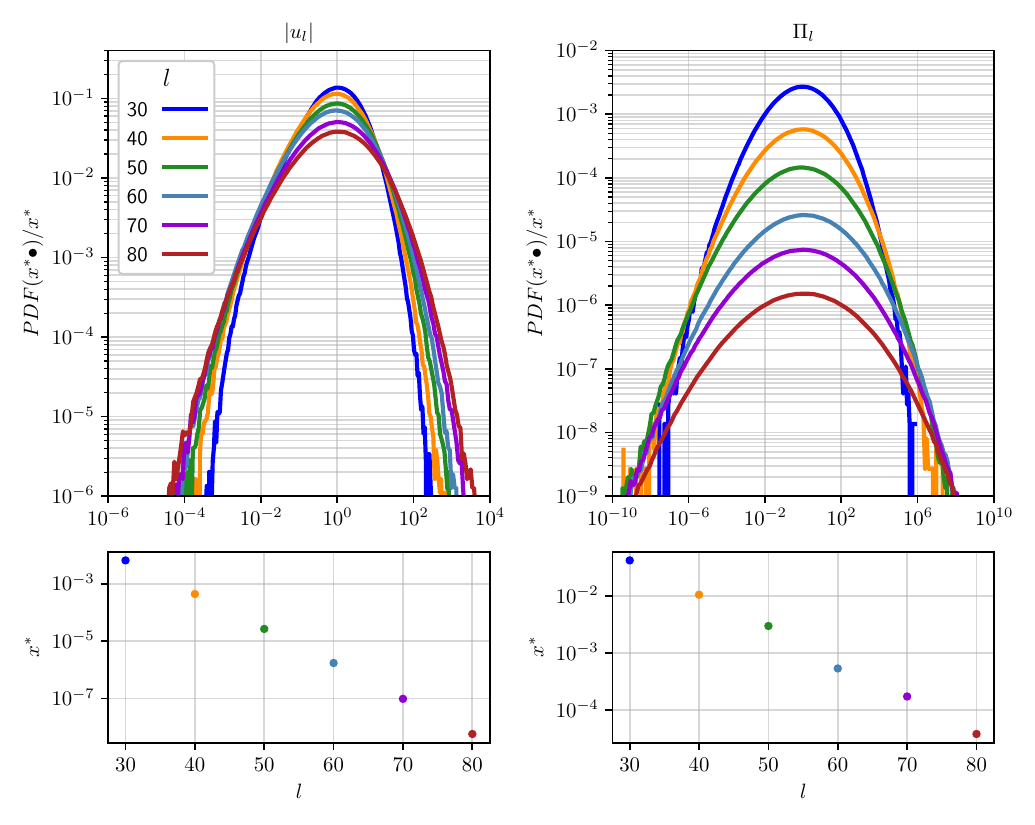}
\caption{Probability distributions of velocity and energy flux, rescaled by their maxima for the sake of visualization. The scaling factors are shown below.}
\label{figPdfs}
\end{figure}

\begin{figure}[t]
\centering
\includegraphics[width=\columnwidth]{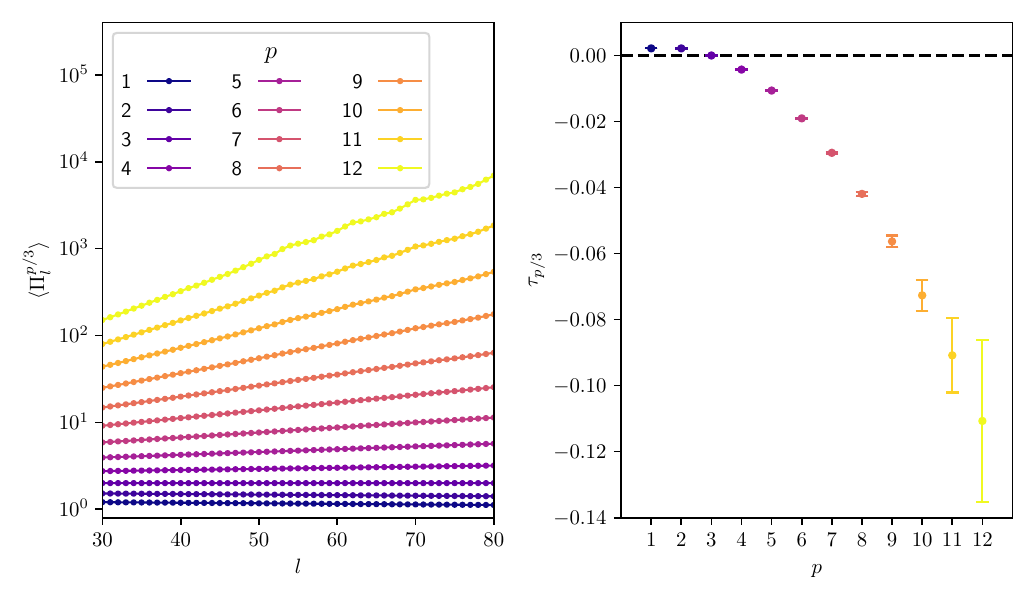}
\caption{Energy-transfer structure functions and corresponding anomalous scaling exponents.}
\label{figExpo}
\end{figure}
Figure \ref{figHist} shows the distributions of $\lvert r\rvert$, $\lvert \gamma\rvert$, $\pi$ and $\lvert \phi\rvert$ for different levels $l$.
The collapse for sufficiently large $l$ verifies the prediction (\ref{ergo}). This collapse is reached very rapidly.
Indeed, after $l=30$, the distributions of amplitude multipliers are indistinguishable. 
The probability densities of amplitude multipliers are monotonically decreasing, whereas the flux multipliers display a uniform distribution. \medskip

In contrast to individual multipliers, their products, represented by $u_{l,n_l}$ and $\Pi_{l,n_l}$, are not self-similar.
This is illustrated in Figure \ref{figPdfs}, which displays their measured distributions. 
The probability densities of $\lvert u\rvert$ and $\Pi$ are linearly rescaled to vertically align the maxima. 
These scaling factors are reported in the panel below. As $l$ increases, stronger tails form, implying the presence of more extreme events. 
\medskip

Figure \ref{figExpo} shows the measured values of the different moments of the energy flux. 
They were obtained from a simulation with $\epsilon=0.4$. A clear power law is observed for all examined moments:
\begin{equation}
    \E\big[ \lvert 2^l\Pi_{l,n}\rvert^{p/3}\big] \propto\lambda^{-l\tau_{p/3}}.
\end{equation}
To compute the exponents, we used the precise error-analysis method proposed in~\cite{deWit2024}, accounting for both statistical and systematic uncertainties. 
The results, including error bars, clearly show that the dimensional prediction corresponding to Kolmogorov scaling $\tau_{p/3}=0$ is violated, thus demonstrating intermittency for this stationary system.
The only exponent without an intermittency correction is $\tau_1=0$, as predicted by (\ref{prediction}). The observed overall behavior is consistent with results obtained for the Navier–Stokes equations \cite{chen2003}. However, we note that the values of $\tau_{p/3}$ depend on the choice of $\epsilon$, with stronger deviations from the Kolmogorov scaling obtained for smaller values of $\epsilon$. Similar results are obtained for the velocity exponents:
\begin{equation}
    \E\big[ \lvert u_{l,n_l} \rvert^p \big] \propto\lambda^{-l\zeta_p}.
\end{equation}
although the error bars are much wider in this case.
\medskip

As a final remark, we note that following sampling methods different from the one described in Section \ref{samp_process} we have also found solutions that violate the connection suggested by dimensional analysis $\zeta_p=\tau_{p/3}+\frac{p}{3}$ between the scaling of the energy flux and that of the cube of the velocity amplitude.
This behavior was not observed in the multi-branch Desnianskii–Novikov model \cite{ajzner2023}. 
In that case, only one random degree of freedom is present—namely, the fraction of energy partitioned between the two children—so that once the intermittency properties of the flux are fixed, the velocity scaling is also determined. Here, instead, three random degrees of freedom are present, but only one controls the energy partition. This stems from the fact that GOY interactions are less local than those in the Desnianskii–Novikov model, resulting in a much larger freedom for the stationary solutions.


\section{Conclusions}

We have introduced a multi-branch shell model with GOY-type couplings and shown that the stationary, inviscid problem admits a concise geometric formulation. In multiplier variables, quartets lie on an affine hyperplane, while physical admissibility (boundedness and forward, positive energy transfer) reduces to the intersection with two half-spaces. This structure yields a transparent constraint geometry for the cascade and underpins the main existence result.\medskip

The geometric viewpoint makes the selection mechanism of turbulence-like cascades explicit: flux constraints delineate a high-dimensional feasible region whose non-trivial intersection guarantees stationary forward-transfer states. In addition, the Markovian sampling along branches provides a principled route to scaling via the spectral theory of positive operators, bridging deterministic constraints and stochastic phenomenology.\medskip

The same construction can be extended to a GOY–Desnianskii–Novikov model, like the one studied in \cite{biferale1994} for the single-branch case, or by introducing horizontal interactions that yield products of multipliers whose number of factors depends on their lowest common ancestor, reflecting the underlying ultrametric structure.
\medskip

We have also demonstrated how intermittency can manifest itself in the multi-branch topology considered here and has allowed us to connect the observed anomalous exponents with the spectrum of the relevant Feynman–Kac operators.
Although such a procedure cannot be directly applied to the Navier–Stokes equations, it nevertheless indicates a promising direction for future investigations.
\medskip

In conclusion, we have proposed and analysed a hierarchical model that is richer than standard single-branch shell models. Furthermore, it reproduces most of the features characteristic of turbulent Navier–Stokes dynamics, most notably the cascade mechanism, while remaining analytically and computationally much simpler than the full Navier–Stokes system.
In this sense, we believe that the present model provides a valuable alternative to the full Navier–Stokes equations for testing theoretical ideas and understanding the mechanisms underlying the cascade. Future work will focus on the dynamics of the model to further assess this statement.



\appendix
\section{Inductive construction}
\label{app1}

We present here the proof of Proposition \ref{thm}, stated in Section \ref{exres}.
\begin{proof}
    {\it Inductive base.} 
    Considering zero boundary conditions for the nonlinear term (\ref{nonlinear}), the first iteration reads
    \begin{equation}
        {r_{1,1}}^2(r_{2,1}+r_{2,2})+{r_{1,2}}^2(r_{2,3}+r_{2,4})=0,
    \end{equation}
    which is incompatible with the positivity of
    \begin{equation}
        \Pi_{1,n}=\frac{a}{4}\,\lambda^{-1}\,{r_{1,n}}^2(r_{2,2n-1}+r_{2,2n})
    \end{equation}
    for both edges. 
    Mathematically, this issue is resolved by following the branch corresponding to the downward flux and discarding the other part of the tree, which is equivalent to imposing a different set of boundary conditions on the nonlinear term. 
    Physically, as already mentioned, we consider a forcing that injects energy at large scales, and the new boundary condition plays the role of this large-scale forcing. 
    The next step is also special because of the reduced form of the parameters,
    \begin{equation}
        \beta=-b\,{r_{1,n}}^{-1}\frac{1}{2}(r_{2,2n-1}+r_{2,2n}),\quad\phi=-\frac{a}{b}\,\beta\neq 0.
    \end{equation}
    The reasoning, however, proceeds analogously to the inductive step.\\
    {\it Inductive step.}
    Let $\boldsymbol{\gamma}(t_3)\vcentcolon=\prj\mathbf{0}+t_3\mathbf{b}_3
    =(\gamma_1,\gamma_1,\gamma_2,\gamma_2)$ for $\sigma t_3\in(L_0,U_0)=I_0$, 
    which is non-empty as proved by (\ref{nonempty}). 
    If $|\gamma_1|<\eta$, we apply the tangential shift $t_1\mathbf{b}_1$ within $A$, 
    with $t_1\in[\eta+|\gamma_1|,\xi-|\gamma_1|]$, which is non-empty provided that
    \begin{equation}
        \xi \ge 3\eta.
    \end{equation}
    The same argument applies if $|\gamma_2|<\eta$.
    Otherwise, boundedness must be enforced. Recall that the conditions can be written as
    \begin{equation}
        \sigma\gamma_1>\sigma\psi_1,\quad\sigma\gamma_2>\sigma\psi_2,\quad-\xi<\sigma\gamma_1<\xi,\quad-\xi<\sigma\gamma_2<\xi.
    \end{equation}
    The values of $\sigma t_3$ that simultaneously satisfy all these conditions form the intersection of three intervals: $I_0$, given by the flux constraints, and $I_1$, $I_2$ from the boundedness requirements. All these intervals are non-empty, open, and connected on the real line, so that a necessary and sufficient condition is that all pairwise intersections be nonempty. Recall that two open connected intervals overlap if and only if the distance between their centers is smaller than the sum of their half-widths.
    We have $I_0=(C_0-\Delta_0,C_0+\Delta_0)$, with
    \begin{align}
        C_0&=\frac{\sigma}{\alpha_2}\bigg[\psi_1-\frac{\beta}{\lVert\boldsymbol{\alpha}\rVert^2}\alpha_1\bigg]+\frac{\lvert\phi\rvert}{4\alpha_1\alpha_2}\\
        &=\frac{\sigma}{\alpha_1}\bigg[\frac{\beta}{\lVert\boldsymbol{\alpha}\rVert^2}\alpha_2-\psi_2\bigg]-\frac{\lvert\phi\rvert}{4\alpha_1\alpha_2},\\
        \Delta_0&=\frac{\lvert\phi\rvert}{4\alpha_1\alpha_2};
    \end{align}
    $I_1=(C_1-\Delta_1,C_1+\Delta_1)$, with
    \begin{align}
        C_1&=-\sigma\frac{\beta}{\lVert\boldsymbol{\alpha}\rVert^2}\frac{\alpha_1}{\alpha_2},\\
        \Delta_1&=\frac{\xi}{\alpha_2};
    \end{align}
    and $I_2=(C_2-\Delta_2,C_2+\Delta_2)$, with
    \begin{align}
        C_2&=\sigma\frac{\beta}{\lVert\boldsymbol{\alpha}\rVert^2}\frac{\alpha_2}{\alpha_1},\\
        \Delta_2&=\frac{\xi}{\alpha_1}.
    \end{align}
    We now evaluate the pairwise intersections. First, $I_1\cap I_2\neq\emptyset$ if and only if
    \begin{equation}
        \lvert C_1-C_2\rvert=\frac{\lvert\beta\rvert}{\lVert\boldsymbol{\alpha}\rVert^2}\bigg(\frac{\alpha_2}{\alpha_1}+\frac{\alpha_1}{\alpha_2}\bigg)<\xi\bigg(\frac{1}{\alpha_1}+\frac{1}{\alpha_2}\bigg)=\Delta_2+\Delta_1
    \end{equation}
    or equivalently,
    \begin{equation}
        \xi>\frac{\lvert\beta\rvert}{2(\alpha_1+\alpha_2)}.
    \end{equation}
    To obtain a uniform bound, we estimate the left-hand side using the inductive hypothesis:
    \begin{equation}
        \frac{\lvert\beta\rvert}{2(\alpha_1+\alpha_2)}\le\frac{\lvert c\rvert+\lvert b\rvert\eta^3}{a\eta^5}
    \end{equation}
    Second, $I_0\cap I_1\neq\emptyset$ if and only if
    \begin{equation}
        \lvert C_0-C_1\rvert=\bigg\lvert\frac{\sigma}{\alpha_2}\psi_1+\frac{\lvert\phi\rvert}{4\alpha_1\alpha_2}\bigg\rvert<\frac{\xi}{\alpha_2}+\frac{\lvert\phi\rvert}{4\alpha_1\alpha_2}=\Delta_1+\Delta_0.
    \end{equation}
    The worst case requires
    \begin{equation}
        \lvert\psi_1\rvert<\xi
    \end{equation}
    and again, estimating under the inductive hypothesis gives
    \begin{equation}
        \xi>\frac{\lvert c\rvert}{a\eta^2}.
    \end{equation}
    The same reasoning applies to the last intersection, $I_0\cap I_2\neq\emptyset$.
    Collecting all requirements yields the expression for the minimal $\xi_0$.
\end{proof}
The same idea readily generalizes to stricter constraints on the flux multipliers,
\begin{equation}
    \pi_{l+1,2n-1}>\epsilon, \qquad \pi_{l+1,2n}>\epsilon,
\end{equation}
with $\epsilon\in(0,1/2)$.
We then obtain the new bounds for $\sigma t_3$:
\begin{equation}
    L_0^{(\epsilon)}=\frac{\sigma}{\alpha_2}\bigg[\psi_1-\frac{\beta}{\lVert\boldsymbol{\alpha}\rVert^2}\alpha_1+\epsilon\frac{\phi}{2\alpha_1}\bigg],\quad
    U_0^{(\epsilon)}=\frac{\sigma}{\alpha_1}\bigg[\frac{\beta}{\lVert\boldsymbol{\alpha}\rVert^2}\alpha_2-\psi_2-\epsilon\frac{\phi}{2\alpha_2}\bigg]
\end{equation}
so that the interval width reads
\begin{equation}
    U_0^{(\epsilon)}-L_0^{(\epsilon)}=\frac{\lvert\phi\rvert}{2\alpha_1\alpha_2}(1-2\epsilon).
\end{equation}


\section*{Acknowledgments}

This work received financial support from the CNRS through the MITI interdisciplinary initiatives, under its exploratory research program. SC acknowledges funding from the ANR SPEED project ANR-20-CE23-0025-01.

\bibliographystyle{elsarticle-num} 
\bibliography{references}

\end{document}